\documentclass{iau}
\usepackage{graphicx}

\title[Magnetic hot supergiants] 
{The evolution of magnetic fields in hot stars\thanks{Based on observations obtained at the Canada-France-Hawaii Telescope (CFHT) operated by the National Research Council of Canada, the Institut National des Sciences de l'Univers of the CNRS of France, and the University of Hawaii, and at the European Southern Observatory (ESO), Chile (program ID 094.D-0274A, 094.D-0274B, 095.D-0155A, 096.D-0072A, and 097.D-0156A).}}

\author[M. E. Oksala et al.]   
{Mary E. Oksala$^{1,2}$, Coralie Neiner$^2$, Cyril Georgy$^3$, Norbert Przybilla$^4$, Zsolt Keszthelyi$^{5,6}$, Gregg Wade$^5$, St\'ephane Mathis$^{7,2}$, Aurore Blaz\`ere$^{8,2}$, Bram Buysschaert$^{2,9}$} 

\affiliation{$^1$ Department of Physics, California Lutheran University, 60 West Olsen Road \#3700, Thousand Oaks, CA 91360, USA\\ email: {\tt moksala@callutheran.edu}\\[\affilskip]
$^2$ LESIA, Observatoire de Paris, PSL Research University, CNRS, Sorbonne Universit\'es, UPMC Univ. Paris 06, Univ. Paris Diderot, Sorbonne Paris Cit\'e, 5 place Jules Janssen, 92195 Meudon, France\\
$^3$ Geneva Observatory, University of Geneva, chemin des Maillettes 51, 1290 Sauverny, Switzerland\\
$^4$ Institut f\"ur Astro- und Teilchenphysik, Universit\"at Innsbruck, Technikerstr. 25/8, 6020, Innsbruck, Austria\\
$^5$ Department of Physics, Royal Military College of Canada, PO Box 17000 Station Forces, Kingston, ON K7K 0C6, Canada\\
$^6$ Department of Physics, Engineering Physics and Astronomy, Queen'€™s University, 99 University Avenue, Kingston, ON K7L 3N6, Canada\\
$^7$ Laboratoire AIM Paris-Saclay, CEA/DRF - CNRS - Universit\'e Paris Diderot, IRFU/SAp Centre de Saclay, 91191 Gif-sur-Yvette, France\\
$^8$ Institut d'Astrophysique et de G\'eophysique, Universit\'e de Li\`ege, Quartier Agora (B5c), All\'ee du 6 ao\^ut 19c, 4000 Sart Tilman, Li\`ege, Belgium\\
$^9$ Instituut voor Sterrenkunde, KU Leuven, Celestijnenlaan 200D, 3001, Leuven, Belgium}
\pubyear{2017}
\volume{329}  
\pagerange{??--??}
\jname{The Lives and Death-Throes of Massive Stars}
\editors{A.C. Editor, B.D. Editor \& C.E. Editor, eds.}
\begin{document}

\maketitle

\begin{abstract}
Over the last decade, tremendous strides have been achieved in our understanding of magnetism in main sequence hot stars. In particular, the statistical occurrence of their surface magnetism has been established ($\sim$10\%) and the field origin is now understood to be fossil. However, fundamental questions remain: how do these fossil fields evolve during the post-main sequence phases, and how do they influence the evolution of hot stars from the main sequence to their ultimate demise? Filling the void of known magnetic evolved hot (OBA) stars, studying the evolution of their fossil magnetic fields along stellar evolution, and understanding the impact of these fields on the angular momentum, rotation, mass loss, and evolution of the star itself, is crucial to answering these questions, with far reaching consequences, in particular for the properties of the precursors of supernovae explosions and stellar remnants. In the framework of the BRITE spectropolarimetric survey and LIFE project, we have discovered the first few magnetic hot supergiants. Their longitudinal surface magnetic field is very weak but their configuration resembles those of main sequence hot stars. We present these first observational results and propose to interpret them at first order in the context of magnetic flux conservation as the radius of the star expands with evolution. We then also consider the possible impact of stellar structure changes along evolution.
\end{abstract}

\firstsection 
\section{Introduction}

Large-scale surveys of hot, massive stars, such as MiMeS (Wade et al. 2016) and BOB (Morel et al. 2015), have recently revealed that $\sim$10\% of hot stars host magnetic fields on the main sequence (Grunhut \& Neiner, 2015; Grunhut et al. 2017, see also Wade et al., these proceedings).  Typically, the structure and strength of these fields are quite homogeneous, with dipole-dominated structure and strength ranging from $\sim$0.3 - 20 kG.
Similar magnetic field structure and strength are seen in massive pre-main sequence (PMS) stars, primarly Herbig Ae/Be stars, indicating that magnetic PMS massive stars are the progenitors of their main sequence (MS) counterparts (Alecian et al. 2013).  Presently lacking, however, is information about the presence of such fields in the post-main sequence stellar evolution phases. 

A realistic picture of what happens to these magnetic fields as the star begins its evolution off the MS and toward the red side of the HR diagram is important to understand not only the impact of the field on the star's evolution, but also to determine how the star's evolution can impact the magnetic field strength and structure. 
The role of magnetism at evolved phases is essential, as it impacts mass loss, stellar angular momentum redistribution and loss, internal mixing of nucleosynthetic products, differential rotation, and convection (e.g., Maeder et al. 2014).
Ultimately, the most massive stars will end their lives as supernovae and understanding the path from the MS to this end point will give insight into the different circumstances that will affect the final state, including the field's influence on the energy budget of the explosion and the final rotation and convection properties.

Despite the wealth of spectropolarimetric data obtained in the last decade, until recently the only two known cases of evolved hot stars with detected magnetic fields were the O9.5I star $\zeta$ Ori Aa, a barely-evolved O-type star located in a binary system with a polar field strength of $\sim$140 G (Blaz\`ere et al. 2015), and the B1.5II star $\epsilon$ CMa, evolved just to the end of the MS and hosting a weak field of a few tens of gauss (Fossati et al. 2015). No strongly evolved magnetic hot stars was known. 

The observed magnetic fields of evolved hot stars are expected to weaken due to magnetic flux conservation, in which an enlarged stellar radius causes a decrease in the surface magnetic field. The surface magnetic field observed should change with time ($t$) according to the relation: 
$B(t) = \frac{B_{MS} R^2_{MS}}{R(t)^2}$,
where the subscript MS indicates the radius ($R$) and polar field strength ($B$) at a particular point on the main sequence. 
If we know the distribution of magnetic field strengths expected for various spectral types on the MS (Shultz et al., these proceedings) and combine this with radius predictions from stellar evolution models, we can roughly predict the secular variation of the measurable magnetic field throughout the star's evolution, if we assume that magnetic flux conservation is the only appreciable effect (Keszthelyi et al., these proceedings).  However, Fossati et al. (2016) indicate that other processes may act to enhance the decline in observed magnetic field strength, acting as a magnetic decay.  Moreover, these evolved hot stars begin to develop subsurface convective regions and dynamo action, that could interact with the fossil magnetic field, causing further changes.

\section{BRITEpol Survey}

The BRITE Constellation of nano-satellites (Weiss et al. 2014) was launched with the goal of time-series photometric observations of every star with magnitude $\rm{V}\leq 4$.  In response to this substantial endeavor, a coinciding magnitude-limited survey was planned to observe each BRITE target using the technique of spectropolarimetry, to deduce the incidence of magnetism across all spectral types.

The BRITE spectropolarimetric (BRITEpol) survey (Neiner et al. 2016) aims to observe $\sim$600 stars with 3 high-resolution spectropolarimetric instruments: HARPSpol (R$\sim$100,000) at the 3.6-m ESO telescope in Chile, Narval (R$\sim$68,000) at T\'el\'escope Bernard Lyot (TBL) in France, and ESPaDOnS (R$\sim$68,000) at the Canada-France-Hawaii Telescope (CFHT) in Hawaii. By observing a single circular polarization spectrum or Stokes~$V$ spectrum, the presence of a magnetic field can be detected, followed up with a second observation for confirmation. The survey detected 52 magnetic stars of a wide variety of spectral types ranging from M-type stars to O-type stars, from PMS stars to evolved stars. In particular, two magnetic evolved hot stars were discovered: $\iota$\,Car and HR\,3890. Currently, follow up monitoring of the most interesting magnetic stars, including all hot stars, is in progress.

\subsection{$\iota$\,Car}

$\iota$\,Car (HR\,3699) is an A7Ib supergiant, for which we determined $T_{\rm{eff}} = 7500\pm150$ K and $\log(g) = 1.85\pm0.1$. Stellar evolution models (Georgy et al. 2013) indicate the star has evolved off the MS, and is currently either in its first crossing of the HR diagram or on the blue loop, but in either case clearly evolved. The star is found to have an age of 19-56 Myr, and an enlarged radius of 50-70 R$_{\odot}$.

The observed Stokes~$V$ spectrum of $\iota$\,Car, when analyzed with the multi-line technique Least-Squares Deconvolution (LSD; Donati et al. 1997), revealed an obvious magnetic signature (Fig.~\ref{fig_stokes}, left panel).  This initial analysis allows us to derive a current dipolar field strength of $B_{\rm{pol}} \geq 3$ G. Based on the star's current and past radius estimates and the equation in Sect.~1, the magnetic dipolar field strength that would have been measurable when the star was still on the MS is $B_{\rm{MS}} \sim350-800$ G, if we consider that the surface field strength is changed purely by magnetic flux conservation. This range is consistent with the typical magnetic field strengths currently observed in massive stars on the MS. Repeated observations of the star over a period of $\sim1.5$ years indicate a rotation period of $\sim2$ years based on the modulation of the magnetic field signature (see Fig.~\ref{fig_stokes}).

\subsection{HR\,3890}

\begin{figure}[t]
\begin{center}
 \includegraphics[width=1.72in]{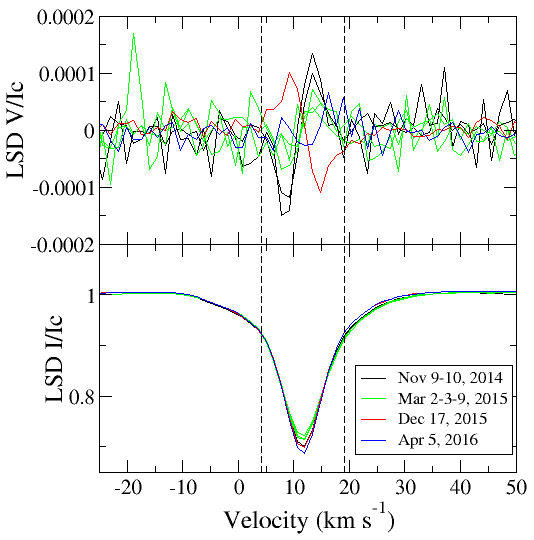} 
 \includegraphics[width=1.72in]{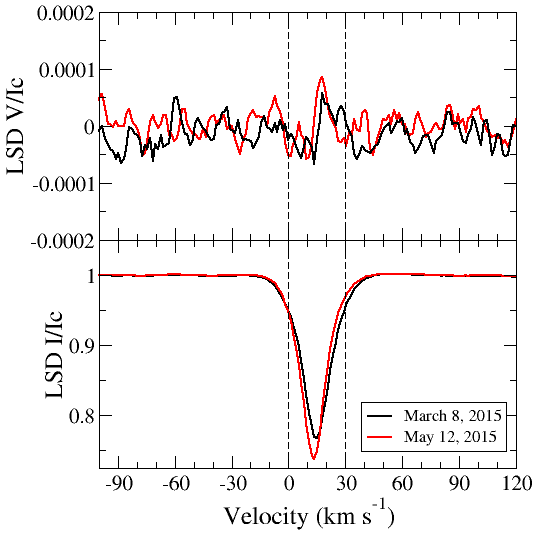} 
 \includegraphics[width=1.72in]{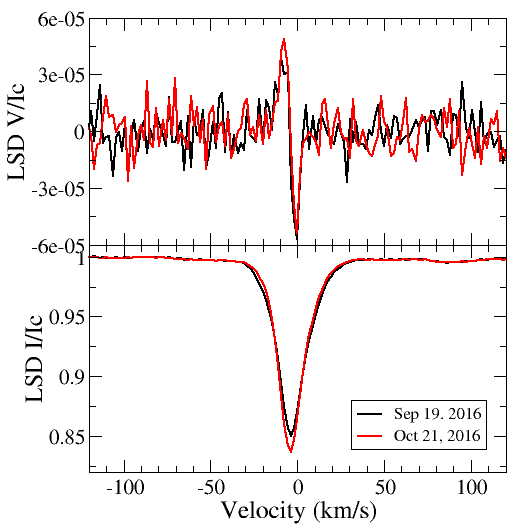} 
 \caption{LSD Stokes I (bottom) and V (top) profiles of the magnetic giant and supergiant stars $\iota$\,Car (A7Ib; left), HR\,3890 (A8Ib; middle), and 19\,Aur (A5II; right).}
   \label{fig_stokes}
\end{center}
\end{figure}

HR\,3890 is an A8Ib supergiant with preliminary parameters determined by this study as $T_{\rm{eff}} = 7500\pm150$ K and $\log(g) = 1.00\pm0.1$. Stellar evolution models paired with these parameters indicate a post-MS evolutionary state, on its initial crossing of the HR diagram. The stellar age is estimated to be between 6-12 Myr, and the star has a radius between 190-300 R$_{\odot}$. The LSD average Stokes~$V$ profile shown in Fig.~\ref{fig_stokes} (middle panel) reveals a magnetic signature that corresponds to a dipolar magnetic field strength of $B_{\rm{pol}} \geq 1$ G.  If we consider again the relationship for the conservation of magnetic flux, and the estimated MS and current radii of HR\,3890, the MS dipolar field strength would be in the range $B_{\rm{MS}} \sim 1450-2150$ G, again consistent with our current known properties of magnetism of MS massive stars.  Additional observations are necessary to identify any variation in the Stokes~$V$ signature.

\section{LIFE Survey}

To fill the void of known magnetic evolved hot stars, study the evolution of their fossil magnetic fields along the stellar evolution sequence, and understand the impact of these magnetic fields on the angular momentum, rotation, mass loss, and evolution of the star itself, we have started a project called LIFE (the Large Impact of magnetic Fields on the Evolution of hot stars) to observe a statistical sample of bright (V$<$8), evolved (luminosity classes I, II, and III), OBA stars with ESPaDOnS at CFHT. We aim to reach very high sensitivity (below 1 G for the longitudinal field measurements) to detect weak fossil fields and possible dynamo fields. The goal is to detect more magnetic evolved hot stars and characterize the properties of the magnetic population, to test the magnetic flux conservation scenario, search for signatures of dynamos, field decay or enhancement, etc.  The observations of 11 LIFE targets so far led to the detection of a first magnetic star: 19\,Aur. 

\subsection{19\,Aur}

19\,Aur (HR\,1740) is an A5II giant, with a reported $T_{\rm{eff}}$ between 6300 and 8600 K and $\log(g)$ between 1.7 and 2.2 (e.g. Lyubimkov et al. 2010, Soubiran et al. 2016). These stellar parameters indicate an estimated $M\sim7$ M$_{\odot}$ and $R\sim35$ R$_{\odot}$. The LSD average Stokes~$V$ profile (Fig~\ref{fig_stokes}, right panel) reveals a clear magnetic signature corresponding to a dipolar magnetic field strength of $B_{\rm{pol}} \geq 6$ G.  Magnetic flux conservation predicts that the MS field strength of 19\,Aur would have been $B_{\rm{MS}} \sim 500$ G, typical of MS stars.

\begin{figure}[t]
\begin{center}
 \includegraphics[width=3.1in]{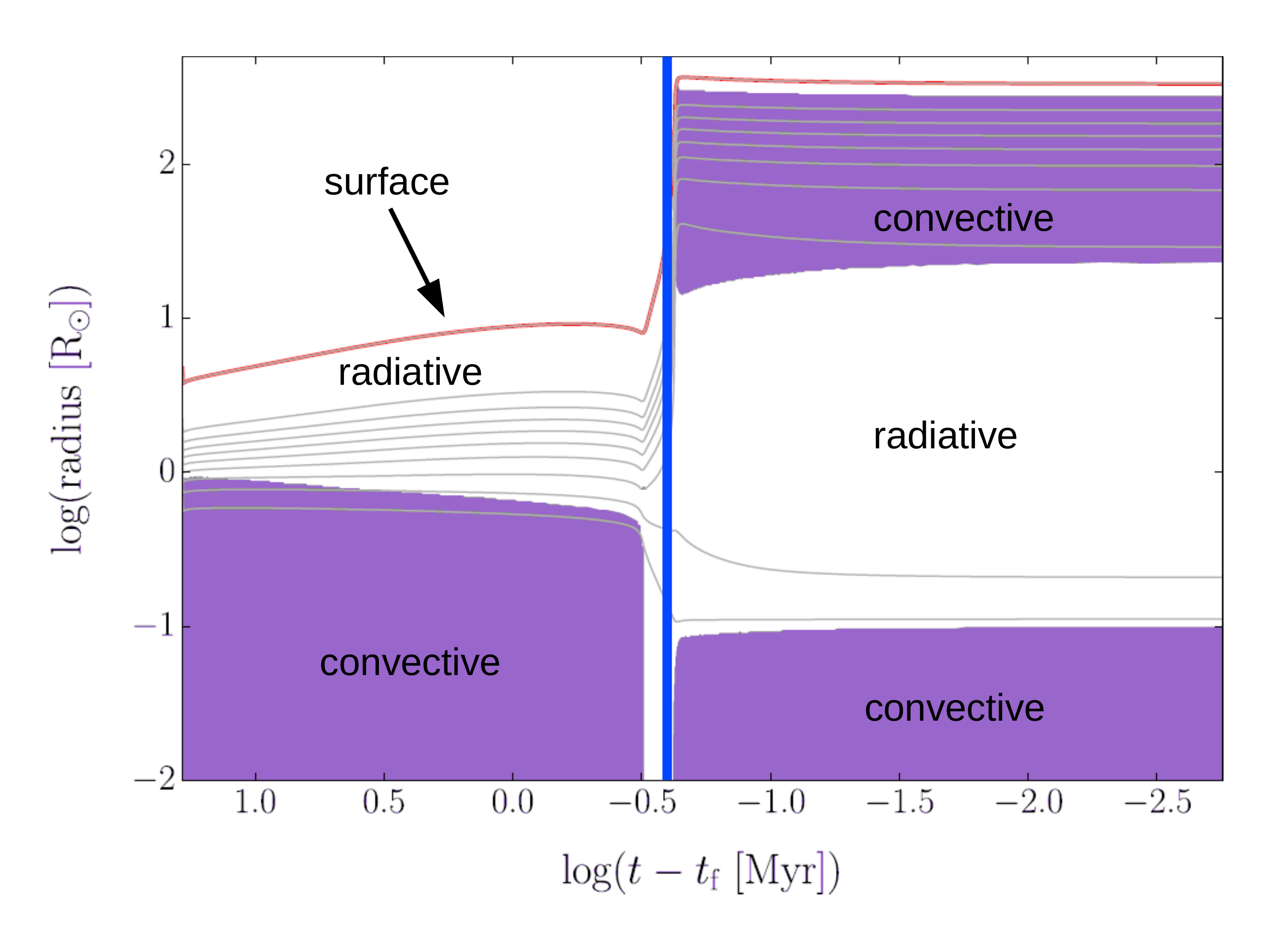} 
 \caption{Kippenhahn diagram showing the evolution of the structure of the magnetic supergiant star $\iota$\,Car. Convective zones are indicated in purple, while radiative zones are in white. The red line indicates the surface of the star and the thin grey lines show the mass distribution. The vertical blue line shows the current position of the star in this diagram. Figure adapted from Neiner et al. (in preparation).}
   \label{fig_kippen}
\end{center}
\end{figure}

\section{Impact of internal structure evolution}

As massive stars evolve, convective regions appear at the top of the radiative zone below the stellar surface (see Fig.~\ref{fig_kippen}). In these regions, dynamo fields could develop (see Augustson et al., these proceedings) and interact with the initial, already present, fossil magnetic field. Using MHD simulations, Featherstone et al. (2009) showed that the interactions between a dynamo and fossil field inside a star may increase the strength of the dynamo and modify the obliquity of the fossil field. Whether  such effects indeed occur will have to be tested once the LIFE project has provided a sufficient number of magnetic evolved hot stars. 

\section{Conclusions}

At present there are only a handful of known magnetic evolved hot stars. Their surface magnetic fields are very weak, often of only a few gauss. At first order, these observed field strengths agree with the basic assumption of magnetic flux conservation during stellar evolution. The LIFE project aims to expand our knowledge of magnetism in evolved hot stars by obtaining high quality observations of a larger number of giants and supergiants to obtain statistical information, test the flux conservation scenario, search for signatures of complex dynamo fields, and identify other potential field decay and enhancement effects.


\begin{thebibliography}{}

\bibitem[Alecian \etal\ (2013)]{Alecian_etal13}
{Alecian}, E., {Wade}, G.~A., {Catala}, C., {Grunhut}, J.~H., {Landstreet}, J.~D., {Bagnulo}, S., {B\"ohm}, T., {Folsom}, C.~P., et al., 2013, MNRAS, 429, 1001

\bibitem[Blazere \etal\ (2015)]{Blazere_etal15}
{Blaz\`ere}, A., {Neiner}, C., {Tkachenko}, A., {Bouret}, J.-C., \& {Rivinius}, T., 2015, A\&A, 582, A110

\bibitem[Donati \etal\ (1997)]{Donati_etal97}
{Donati}, J.-F., {Semel}, M., {Carter}, B.~D., {Rees}, D.~E., \& {Collier Cameron}, A., 1997, MNRAS, 291, 658

\bibitem[Featherstone \etal\ (2009)]{Featherstone_etal09}
{Featherstone}, N.~A., {Browning}, M.~K., {Brun}, A.~S., {Toomre}, J., 2009, ApJ, 705, 1000

\bibitem[Fossati \etal\ (2015)]{Fossati_etal15}
{Fossati}, L., {Castro}, N., {Morel}, T., {Langer}, N., {Briquet}, M., {Carroll}, T.~A., {Hubrig}, S., {Nieva}, M.~F., et al., 2015, A\&A, 574, A20

\bibitem[Fossati \etal\ (2016)]{Fossati_etal16}
{Fossati}, L., {Schneider}, F.~R.~N., {Castro}, N., {Langer}, N., {Sim\'on-D{\'{\i}}az}, S., {M\"uller}, A., {de Koter}, A., {Morel}, T., et al., 2016, A\&A, 592, A84

\bibitem[Georgy \etal\ (2013)]{Georgy_etal13}
{Georgy}, C., {Ekstr\"om}, S., {Granada}, A., {Meynet}, G., {Mowlavi}, N., {Eggenberger}, P., \& {Maeder}, A., 2013, A\&A, 553, A24

\bibitem[Grunhut \& Neiner (2015)]{GN_15}
{Grunhut}, J.~H. \& {Neiner}, C., 2015, in K.~N. {Nagendra}, S. {Bagnulo}, R. {Centeno}, \& M. {Jes\'us Mart{\'{\i}}nez Gonz\'alez} (eds.), IAU Symposium, Vol. 305, pp 53-60

\bibitem[Grunhut \etal\ (2017)]{Grunhut_etal17}
{Grunhut}, J.~H., {Wade}, G.~A., {Neiner}, C., {Oksala}, M.~E., {Petit}, V., {Alecian}, E., {Bohlender}, D.~A., {Bouret}, J.-C., et al., 2017, MNRAS, 465, 2432

\bibitem[Lyubimkov \etal\ (2010)]{Lyub_etal10}
{Lyubimkov}, L.~S., {Lambert}, D.~L., {Rostopchin}, S.~I., {Rachkovskaya}, T.~M., \& {Poklad}, D.~B., 2010, MNRAS, 402, 1369

\bibitem[Maeder \etal\ (2014)]{Maeder_etal14}
{Maeder}, A., {Przybilla}, N., {Nieva}, M.-F., {Georgy}, C., {Meynet}, G., {Ekstr\"om}, S., \& {Eggenberger}, P., 2014, A\&A, 565, A39

\bibitem[Morel \etal\ (2015)]{Morel_etal15}
{Morel}, T., {Castro}, N., {Fossati}, L., {Hubrig}, S., {Langer}, N., {Przybilla}, N., {Sch\"oller}, M., {Carroll}, T., et al., 2015, in G. {Meynet}, C. {Georgy}, J. {Groh}, \& P. {Stee} (eds.), IAU Symposium, Vol. 307, pp 342-347

\bibitem[Neiner \etal\ (2016)]{Neiner_etal16}
{Neiner}, C., {Wade}, G., {Marsden}, S., \& {Blaz\`ere}, A., 2016, arXiv:1611.03285

\bibitem[Soubiran \etal\ (2016)]{Soubiran_etal16}
{Soubiran}, C., {Le Campion}, J.-F., {Brouillet}, N., {Chemin}, L., 2016, A\&A, 591, A118

\bibitem[Wade \etal\ (2016)]{Wade_etal16}
{Wade}, G.~A. and {Neiner}, C. and {Alecian}, E. and {Grunhut}, J.~H. and {Petit}, V. and {Batz}, B.~d. and {Bohlender}, D.~A. and {Cohen}, D.~H., et al., 2016, MNRAS, 456, 2

\bibitem[Weiss \etal\ (2014)]{Weiss_etal14}
{Weiss}, W.~W., {Moffat}, A.~F.~J., {Schwarzenberg-Czerny}, A., {Koudelka}, O.~F., {Grant}, C.~C., {Zee}, R.~E., {Kuschnig}, R., {Mochnacki}, S. , et al., 2014, in  {Guzik}, W.~J. {Chaplin}, G. {Handler}, \& A. {Pigulski} (eds.),  IAU Symposium, Vol. 301, pp 67-68 

\end{thebibliography}
\end{document}